# Convolutional sparse coding for capturing high speed video content

Ana Serrano[1]    Elena Garces[1]    Diego Gutierrez[1]    Belen Masia[1,2]

[1] Universidad de Zaragoza - I3A    [2] MPI Informatik

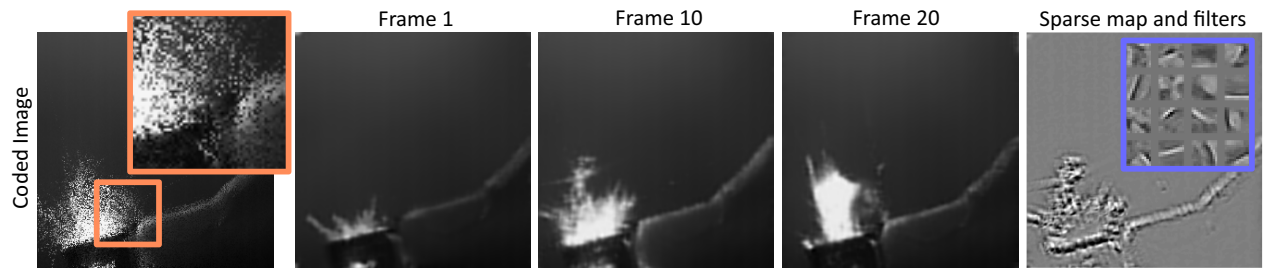

**Figure 1:** *Reconstruction of a high speed video sequence from a single, temporally-coded image using convolutional sparse coding (CSC). The sequence shows a lighter igniting.* Left: *Coded image, from which 20 individual frames will be reconstructed; inset shows a close-up of the coded temporal information.* Middle: *Three frames of the reconstructed video.* Right: *CSC models the signal of interest as a convolution between sparse feature maps and trained filter banks: The image shows a sparse feature map for one of the frames, and the inset marked in blue some of the trained filters.*

**Abstract**

*Video capture is limited by the trade-off between spatial and temporal resolution: when capturing videos of high temporal resolution, the spatial resolution decreases due to bandwidth limitations in the capture system. Achieving both high spatial and temporal resolution is only possible with highly specialized and very expensive hardware, and even then the same basic trade-off remains. The recent introduction of compressive sensing and sparse reconstruction techniques allows for the capture of single-shot high-speed video, by coding the temporal information in a single frame, and then reconstructing the full video sequence from this single coded image and a trained dictionary of image patches. In this paper, we first analyze this approach, and find insights that help improve the quality of the reconstructed videos. We then introduce a novel technique, based on convolutional sparse coding (CSC), and show how it outperforms the state-of-the-art, patch-based approach in terms of flexibility and efficiency, due to the convolutional nature of its filter banks. The key idea for CSC high-speed video acquisition is extending the basic formulation by imposing an additional constraint in the temporal dimension, which enforces sparsity of the first-order derivatives over time.*

Categories and Subject Descriptors (according to ACM CCS): I.4.1 [Computer Graphics]: Digitization and Image Capture—Sampling

## 1. Introduction

During the last years, video capture technologies have seen large progress, due to the necessity of acquiring information at high temporal and spatial resolution. However, cameras still face a basic bandwidth limitation, which poses an intrinsic trade-off between the temporal and spatial dimensions. This trade-off is mainly determined by hardware restrictions, such as readout and analog-to-digital conversion times of the sensors. This makes capturing high speed video at high spatial resolutions *simultaneously* still an open problem.

Recent works try to overcome these limitations either with hardware-based approaches such as the camera array prototype pro-posed by Willburn et al. [WJV*04], or with software-based approaches, like the work of Gupta et al. [GBD*09], where the authors proposed combining low resolution videos with a few key frames at high resolution. Other approaches rely on computational imaging techniques, combining optical elements and processing algorithms [WILH11, MWDG13, WLGH12]. For instance, a novel approach based on the emerging field of compressive sensing was presented [LGH*13, HGG*11]. This technique allows to fully recover a signal even when sampled at rates lower than the Nyquist-Shannon theorem, provided that the signal is sufficiently sparse in a given domain. These works rely on this technique to selectively sample pixels at different time instants, thus coding the temporal information of a frame sequence in a single image. They then recover the full frame sequence from that coded image. The key assumption is that the time varying appearance of the captured scenes can be represented as a sparse linear combination of elements of an overcomplete basis (dictionary). This representation, and the subsequent reconstruction, is done in a patch-based manner, which is not free of limitations. Training and reconstruction usually takes a long time; moreover, the overcomplete basis needed for reconstruction has to be made up of atoms of similar nature to the video that is being reconstructed. This in turn imposes the need to use a specialized, expensive camera to capture such basis.

This paper represents an extended version of our previous work [SGM15], where we i) performed an in-depth analysis of the main parameters defined in Liu's *patch-based* compressive sensing and sparse reconstruction framework [LGH*13]; ii) introduced the Lars/lasso algorithm for training and reconstruction, and showed how it improved the quality of the results; iii) presented a novel algorithm for choosing the training blocks, which further improved performance as well as reconstruction time; and iv), we further explored the existence of a good predictor of the quality of the reconstructed video. These contributions are now briefly summarized in Sections 4, 6, and 7, and in the supplemental material.

In this work, we additionally introduce a novel *convolutional sparse coding* (CSC) approach for high-speed video acquisition, and show how this outperforms existing patch-based sparse reconstruction techniques in several aspects. In particular, both training and reconstruction times are significantly reduced, while the convolutional nature of the atoms allows for reconstruction of videos using generic, content-agnostic dictionaries. This is due to the fact that the basis learned no longer needs to be able to reconstruct each signal block in isolation, instead allowing shiftable basis functions to discover a lower rank structure. In other words, image patches are no longer considered independent; interactions are modeled as convolutions, which translates into a more expressive basis which better reconstructs the underlying mechanics of the signal [BL14]. This completely removes the need to capture similar scenes using an expensive high-speed camera, and to explicitly train a dictionary, which significantly extends the applicability of our CSC framework. As an example, all the results shown in this paper have been reconstructed using an existing dictionary containing *images of fruits* [HHW15]. Note that we sample less than 15% of the pixels, which are additionally integrated over time to form a single image; despite this extremely suboptimal input data, we are able to successfully reconstruct high-speed videos of good quality. We provide source high speed videos, code, and results of our implementation[†].

## 2. Related Work

**Coded exposures.** Coded exposure techniques have been used to improve certain aspects of image and video acquisition in the field of computational photography. The goal is to optically code the incoming light before it reaches the sensor, either with coded apertures or shutter functions. For instance, Raskar et al. [RAT06] proposed the use of a *flutter shutter* to recover motion-blurred details in images. With the same purpose Gu et al. [GHMN10] propose the *coded rolling shutter* as an improvement over the conventional *rolling shutter*. Alternatively, codes in the spatial domain have been used for light field reconstruction [VRA*07], high-dynamic range imaging [NM00], to recover from defocus blur [MCPG11, MPCG12], or to obtain depth information [LFDF07, ZLN09].

**Compressive sensing.** The theory of compressive sensing has raised interest in the research community since its formalization in the seminal works of Candes et al. [CRT06] and Donoho [Don06]. Numerous recent works have been devoted to applying this theory to several fields, including image and video acquisition. In one of the most significant works, the *Single Pixel Camera* of Wakin et al. [WLD*06], the authors introduce a camera prototype with only one pixel, which allows the reconstruction of complete images acquired with several captures under different exposure patterns. Other examples in imaging include the work of Marwah et al. [MWBR13], in which they achieve *light field* acquisition from a single coded image; high dynamic range imaging [SBN*12]; or capturing hyperspectral information [LLWD14, JCK16]. Recently, compressive sensing was proposed to reconstruct high-speed video from a single image [LGH*13, HGG*11], combining coded exposure and dictionary learning. Some of the key design choices and parameters were analyzed in [SGM15], leading to an improvement of the original design.

**Convolutional sparse coding.** Convolutional sparse coding has quickly become one of the most powerful tools in machine learning, with many applications in signal processing, computer vision, or computational imaging, to name a few. Grosse et al. [GRKN07] introduced convolutional constraints to sparse coding, as well as an efficient minimization algorithm to make CSC practical, for the particular problem of 1D audio signals. Since then, many other works have extended this basic, canonical framework, with the goal of making it faster or more efficient (e.g. [BEL13, KF14, HHW15]). Bristow and Lucey [BL14] recently presented a large collection of examples covering different application domains, showing the fast spread and general applicability of CSC. Some examples of application domains include learning hierarchical image representations for applications in vision [CPS*13, SKL10]; decomposition of transient light transport [HDL*14]; imaging in scattering media [HST*14]; or high dynamic range capture [SHG*16]. In this paper we present a practical application of CSC for the particular case of high-speed video acquisition.

---

[†] http://webdiis.unizar.es/~aserrano/projects/CSC-Video.html

## 3. Background on sparse reconstruction

Compressive sensing has revolutionized the field of signal processing by providing a means to reconstruct signals that have been sampled at rates lower than what the Nyquist-Shannon theorem dictates. In general, this undersampling or incomplete acquisition of the signal is represented as:

$$\mathbf{y} = \mathbf{\Phi x} \qquad (1)$$

where the signal of interest (in our case a video sequence) is represented by $\mathbf{x} \in \mathbb{R}^m$, the captured signal (a coded image) is $\mathbf{y} \in \mathbb{R}^n$, with $n \ll m$, and $\mathbf{\Phi} \in \mathbb{R}^{n \times m}$ contains the sampling pattern and is called *measurement matrix*. An example of this sampling in the case of video acquisition is shown in Figure 2. Recovering the signal of interest having as input $\mathbf{\Phi}$ and $\mathbf{y}$ requires solving an underdetermined system of equations and finding a basis in which the signal of interest is sparse, and is known as the *sparse reconstruction* problem.

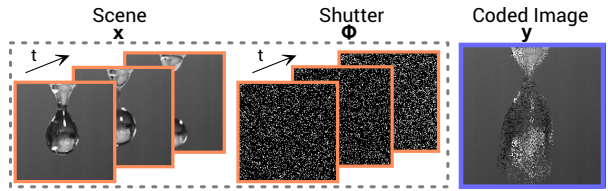

**Figure 2:** *Video acquisition via compressive sensing. The signal of interest $\mathbf{x}$ is sampled using a certain pattern $\mathbf{\Phi}$ which varies in time, and integrated to a single 2D image $\mathbf{y}$. Sparse reconstruction aims to recover $\mathbf{x}$ from $\mathbf{y}$, knowing the sampling pattern $\mathbf{\Phi}$, a severely underdetermined problem. Note that the $\mathbf{\Phi}$ displayed in the image is an actual example of a pattern used for our captures.*

**Patch-based sparse coding.** To solve the sparse reconstruction problem posed by Eq. 1, the signal of interest $\mathbf{x}$ needs to be sparse, meaning that it can be represented in some alternative domain with only a few coefficients. This can be expressed as:

$$\mathbf{x} = \sum_{i=1}^{N} \mathbf{\psi}_i \alpha_i \qquad (2)$$

where $\mathbf{\psi}_i$ are the elements of the basis that form the alternative domain, and $\alpha_i$ are the coefficients, which are in their majority zero or close to zero if the signal is sparse. Many natural signals, such are images or audio, can be considered sparse if represented in an adequate domain.

In order to reconstruct the original signal from the undersampled, acquired one, we jointly consider the sampling process (Eq. 1) together with the representation in the sparse dictionary (Eq. 2), yielding the following formulation:

$$\mathbf{y} = \mathbf{\Phi x} = \mathbf{\Phi \Psi \alpha} \qquad (3)$$

where $\mathbf{\Psi} \in \mathbb{R}^{m \times q}$ represents an overcomplete basis (also called *dictionary*) with $q$ elements. If the original sequence $\mathbf{x}$ is $s-sparse$ in the domain of the basis formed by the *measurement matrix* $\mathbf{\Phi}$ and the dictionary $\mathbf{\Psi}$, it can be well represented by a linear combination of at most $s$ coefficients in $\mathbf{\alpha} \in \mathbb{R}^q$. Note that we are looking for a sparse solution; therefore, the search of the coefficients $\mathbf{\alpha}$ has to be posed as a minimization problem. This optimization will search for the unknown $\mathbf{\alpha}$ coefficients, seeking a sparse solution to Eq. 3. This is typically formulated in terms of the $L_1$ norm, since $L_2$ does not provide sparsity and $L_0$ presents an ill-posed problem which is difficult to solve:

$$\min_{\mathbf{\alpha}} \|\mathbf{\alpha}\|_1 \; subject \; to \; \|\mathbf{y} - \mathbf{\Phi \Psi \alpha}\|_2^2 \leq \varepsilon \qquad (4)$$

where $\varepsilon$ is the residual error. Eq. 4 is usually solved in a patch-based manner, dividing the signal spatially into a series of blocks (in the case of videos, the blocks are of size $p_x \times p_y \times p_z$), reconstructing them individually, and merging them to yield the final reconstructed signal.

**Convolutional sparse coding.** As an alternative, recent works propose modeling the $\mathbf{x}$ as a sum of sparsely-distributed convolutional features [GRKN07]:

$$\mathbf{x} = \sum_{k=1}^{K} \mathbf{d}_k * \mathbf{z}_k \qquad (5)$$

where $\mathbf{d}_k$ are a set of convolutional filters that conform the dictionary, and $\mathbf{z}_k$ are sparse feature maps. The filters have a fixed spatial support, and the feature maps are of the size of the signal of interest.

Heide and colleagues [HHW15] presented a formulation for the recovery of a signal $\mathbf{x}$, modeled as in Eq. 5, from a degraded signal $\mathbf{y}$ measured as shown in Eq. 1:

$$\operatorname*{argmin}_{\mathbf{z}} \frac{1}{2} \|\mathbf{y} - \mathbf{\Phi} \sum_{k=1}^{K} \mathbf{d}_k * \mathbf{z}_k\|_2^2 + \beta \sum_{k=1}^{K} \|\mathbf{z}_k\|_1 \qquad (6)$$

In their paper, the authors also propose efficient algorithms to train the filter bank and to solve the minimization problem. Note that training the filter bank amounts to solving the minimization in Eq. 6 optimizing for both the feature maps $\mathbf{z}_k$ and the filters $\mathbf{d}_k$.

**Coded images from high speed video sequences.** In the case of video, the measurement matrix introduced in $\mathbf{\Phi}$ is implemented as a shutter function that samples different time instants for every pixel. The final image is thus formed as the integral of the light arriving to the sensor for all the temporal instants sampled with the shutter function:

$$I(x,y) = \sum_{t=1}^{T} S(x,y,t) X(x,y,t) \qquad (7)$$

where $I(x,y)$ is the captured image, $S$ the shutter function and $X$ the original scene. In a conventional capture system $S(x,y,t) = 1 \; \forall \; x,y,t$ but in this case $S$ should be such that it fulfils the mathematical properties of a measurement matrix suitable for sparse reconstruction, as well as the constraints imposed by the hardware. An easy way to fulfil the mathematical requirements is to build a random sampling matrix. However, since a fully-random sampling matrix cannot be implemented in current hardware, we use the shutter function proposed by Liu et al. [LGH*13], which can be easily implemented in a *DMD (Digital micromirror device)* or an *LCoS (Liquid Crystal on Silicon)* placed before the sensor, and approximates randomness while imposing additional restrictions to make a hardware implementation possible. In particular, the proposed shutter is implemented in a *LCoS*. Each pixel is only sampled once throughout the sequence, with a fixed bump length. We refer the reader to the work of Liu et al. [LGH*13] for more details about the coded shutter implementation.

In Section 4 we will show how to apply traditional, patch-based sparse coding to the problem of high speed video acquisition, and provide an analysis of the most important parameters. Section 5 will then offer an alternative solution, focused on efficiency, by modifying the convolutional sparse coding formulation in Eq. 6. This alternative solution also lifts some of the restrictions imposed by the patch-based sparse coding approach, such as the need to build a dictionary whose atoms are similar to the videos that are going to be reconstructed.

## 4. Patch-based sparse coding approach

This section describes the specifics of how to solve the sparse reconstruction problem using a patch-based approach (described in Section 3) for high speed video. The mathematical formulation is given by Eqs. 2 to 4; here we describe how we train the dictionary $\Psi$, and how we solve the optimization problem in Eq. 4. We summarize the main ideas here, and provide more details in the supplemental material.

**Learning high speed video dictionaries.** We have captured a database of high-speed videos which we use for training and validation of our techniques. The database consists of 14 videos captured at captured at 1000 frames per second with a high-speed Photron SA2 camera, part of which are used for training, and part for testing. The Photron SA2 provides up to 4 Megapixels at a rate of 1000 fps. The acquisition setup is shown in Figure 3. Our database provides scenes of different nature, and a wide variety of spatial and temporal features. Representative frames of the videos in the database are shown in the supplemental material. We learn the fundamental building blocks (atoms) from our captured videos, and create an overcomplete dictionary. For training, we use the DLMRI-Lab implementation [RB11] of the K-SVD [AEB06] algorithm, which has been widely used in the compressive sensing literature. We propose an alternative to random selection that maximizes the presence of blocks with relevant information, by giving higher priority to blocks with high variance.

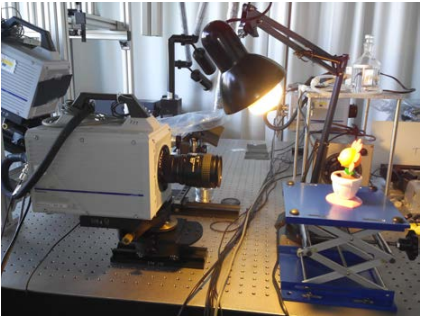

**Figure 3:** *Setup for the acquisition of our high speed video database with a Photron SA2 camera. In order to capture videos with high quality the scene must be illuminated with a strong, direct light.*

**Reconstructing high speed videos.** Once the dictionary $\Psi$ is trained, and knowing the measurement matrix $\Phi$, we need to solve Eq. 4 to estimate the $\alpha$ coefficients and reconstruct the signal. Many algorithms have been developed for solving this minimization problem for compressive sensing reconstruction. We use the implementations available in the SPArse Modeling Software (SPAMS) [MBPS10].

## 5. Introducing CSC for high-speed video acquisition

The use of *patch-based, conventional* sparse coding in the recovery of high-speed video produces good results, as we will show in Section 7. However, *convolutional* sparse coding (CSC) can offer significant improvements: First, learning convolutional filters allows for a richer representation of the signal, since they span a larger range of orientations and are spatially-invariant, as opposed to the patches learnt in a conventional dictionary. Second, due to their convolutional nature, dictionaries made up of filter banks are more versatile, in the sense that they are content-agnostic: they do not need to contain atoms of similar nature to the signals that are to be reconstructed with them. Finally, reconstruction time is a major bottleneck in patch-based approaches, but is significantly reduced in a convolutional framework; this is especially important when dealing with video content.

Recent efficient solutions for CSC have been proposed for images [HHW15, SHG*16]. In principle, adapting these solutions to video could be done simply by extending them to three dimensions (x-y-t), with 3D filters $\mathbf{d}_k$ and 3D feature maps $\mathbf{z}_k$. The optimization in Eq. 6 could then be solved in a manner analogous to 2D [HHW15, Algorithm 1]. Doing this, however, is tremendously computationally expensive[‡]. We therefore revert to a 2D formulation instead, which is computationally manageable, and adapt it to be able to properly deal with video content, as explained next.

In our proposed formulation, the sparse feature maps $\mathbf{z}_k$ and the filters $\mathbf{d}_k$ remain two-dimensional; thus their convolution yields two-dimensional images, which correspond to the individual frames. This per-frame reconstruction of the video could be achieved using the optimization in Eq. 6. To adapt this solution to video, we impose an additional constraint in the temporal dimension, which enforces sparsity of the first-order derivatives over time. The resulting reconstruction process is then:

$$\operatorname*{argmin}_{\mathbf{z}} \frac{1}{2}\beta_d \|\mathbf{y} - \Phi \sum_{k=1}^{K} \mathbf{d}_k * \mathbf{z}_k\|_2^2 + \\ \beta_1 \sum_{k=1}^{K} \|\mathbf{z}_k\|_1 + \beta_2 \sum_{k=1}^{K} \|\nabla_t \mathbf{z}_k\|_1 \quad (8)$$

The operator $\nabla_t$ represents the first order backward finite difference along the temporal dimension: $\nabla_t \mathbf{z}_k = (\mathbf{z}_k^t - \mathbf{z}_k^{t-1}) \, \forall t \in \{2,\ldots,T\}$,

---

[‡] To give a practical example: 32GB of RAM allow training a maximum of *100* filters of size $11 \times 11 \times 10$; while this number of filters is adequate for 2D images, it is insufficient in the presence of the extra dimension in video.

where $T$ is the number of frames being reconstructed. $\beta_d$, $\beta_1$ and $\beta_2$ are the relative weights of the data term, the sparsity term, and the temporal smoothness term, respectively.

A modified ADMM algorithm can be used to solve this problem by posing the objective as a sum of closed, convex functions $f_j$ as follows:

$$\operatorname*{argmin}_{\mathbf{z}} \sum_{j=1}^{J} f_j(\mathbf{K}_j \mathbf{z}) \qquad (9)$$

where $J = 3$, $f_1(\xi) = \frac{1}{2}\beta_d \|\mathbf{y} - \mathbf{\Phi}\xi\|_2^2$, $f_2(\xi) = \beta_1 \|\xi\|_1$, and $f_3(\xi) = \beta_2 \|\xi\|_1$. Consequently, the matrices $\mathbf{K}_j$ are $\mathbf{K}_1 = \mathbf{D}$, $\mathbf{K}_2 = \mathbb{I}$, and $\mathbf{K}_3 = \nabla$, where $\mathbf{D}$ is formed by the convolution matrices corresponding to the filters $\mathbf{d}_k$, and $\nabla$ is the matrix corresponding to the first order backward differences in the temporal dimension, as explained above. We use the implementation of ADMM (*Alternating Direction Method of Multipliers*) described in Algorithm 1 to solve Eq. 9. More details about this implementation can be found in the literature [AF13, HHW15, Algorithm 1].

**Algorithm 1**
1: **for** $i = 1$ to $I$ **do**
2:   $\mathbf{y}^{i+1} = \operatorname*{argmin}_{\mathbf{y}} \|\mathbf{Ky} - \mathbf{z} + \lambda^i\|_2^2$
3:   $\mathbf{z}_j^{i+1} = \mathbf{prox}_{\frac{f_j}{\rho}}(\mathbf{K}_j \mathbf{y}_j^{i+1} + \lambda_j^k) \; \forall j \in \{1, \ldots, J\}$
4:   $\lambda^{i+1} = \lambda^i + (\mathbf{Ky}^{i+1} - \mathbf{z}^{i+1})$
5: **end for**

## 6. Analysis

In this section we discuss implementation details of the system, and we perform an analysis of the two approaches proposed in Section 4 and Section 5, exploring parameters of influence for both methods.

We now analyze the influence of the key parameters for both approaches: patch-based and convolutional sparse coding, and find the parameter combination yielding the best results. None of the test videos were used during training. As measures of quality, we use PSNR (Peak Signal to Noise Ratio), widely used in the signal processing literature, and the MS-SSIM metric [WSB03], which takes into account visual perception. The complete analysis can be found in the supplementary material.

### 6.1. Patch-based sparse coding

**Implementation details**: We use the K-SVD algorithm [AEB06] to train our dictionary and the LARS-Lasso solver [EHJT04] for solving the minimization problem. In order to achieve faithful reconstructions it is important that each atom (patch) is large enough to contain significant features (such as edges or corners), but not too large for avoiding learning very specific features of the training videos (and thus overfitting). We have tested several patch sizes (results included in the supplemental material) and we have chosen the size yielding better quality in the results: $7\,pixels \times 7\,pixels \times 20\,frames$.

**Training a dictionary**: The amount of blocks (3D patches) resulting from splitting the training videos is unmanageable for the training algorithm; thus the dimensionality of the training set has to be reduced. The straightforward solution is to randomly choose a manageable amount of blocks. However, a high percentage of these do not contain meaningful information about the scene (as in static backgrounds). We thus explore several other ways to select the blocks for training:

- **Variance sampling**: We calculate the variance for each block and bin them in three categories: high, medium and low variance. Then we randomly select the same amount of blocks for every bin to ensure the presence of high variance blocks in the resulting set.
- **Stratified gamma sampling**: We sort the blocks by increasing variance and sample them with a gamma curve ($f(x) = x^\gamma$). We analyze the effect of $\gamma = 0.7$, which yields a curve closer to a linear sampling, and $\gamma = 0.3$. The goal of this stratification is to ensure the presence of all the strata in the final distribution. We divide the range uniformly and calculate thresholds for the strata applying the gamma function. Then we randomly choose a sample from every strata and remove that sample from the original set. This process is iterated until the number of desired samples is reached.
- **Gamma sampling**: We choose directly samples from the original set following a gamma curve sampling. We also test values of $\gamma = 0.7$ and $\gamma = 0.3$.

Figure 4 shows results for one of the tested videos, with dictionaries built with the different selection methods explained above. Results for all the videos tested were consistent. *Random* and *Variance sampling* clearly outperform the other methods, with the *Variance sampling* yielding slightly better results.

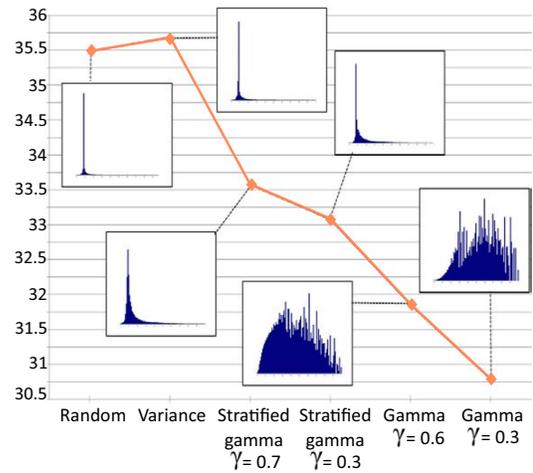

**Figure 4:** *Quality of the reconstruction (in terms of PSNR) for a sample video (*PourSoda*) as a function of the method used to select training blocks for learning the dictionary. For each method we show an inset with the histogram of variances of video blocks of the resulting training set.*

## 6.2. Convolutional sparse coding

**Implementation details**: We use a trained dictionary of 100 filters of size $11 \times 11$ pixels. We have performed tests with different filter sizes, and we have found the reconstruction quality to be very similar between different sizes (see the supplemental material for results). The convolutional nature of the algorithm makes it more flexible, and unlike the patch-based approach, more robust towards variations in the filters size. Nevertheless, we chose to train the filters with a size of $11 \times 11$ pixels because they yield slightly better results. Regarding the amount of filters, we found that 100 filters are enough to make the algorithm converge in the training.

**Training a dictionary**: One of the theoretical advantages of convolutional sparse coding over the patch-based approach is that the learned dictionaries need not be of similar nature to the signal being reconstructed. To analyze this, we compare the quality of the reconstruction for two different dictionaries: A generic dictionary obtained from the *fruits* dataset (it simply contains pictures of fruits) provided by Heide et al. [HHW15], and a specific dictionary trained with a set of frames from our captured video database (different from the ones we then reconstruct). Figure 5 shows that the qual-ity of the reconstructions using both dictionaries is very similar for all the videos analyzed. This confirms the theory for the particular case of high-speed video reconstruction; as a consequence, we no longer need to acquire specific data to train dictionaries, adapted to every particular problem. We use this generic *fruit* dictionary for all our results.

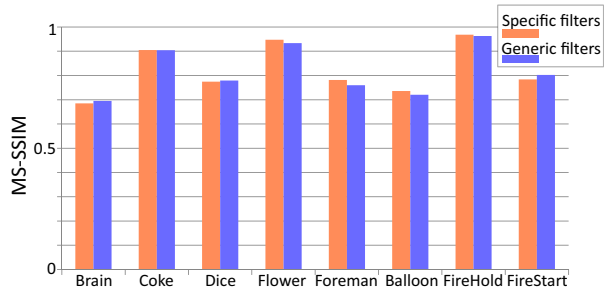

**Figure 5:** *Quality of the reconstruction (in terms of the quality metric MS-SSIM) for two different dictionaries of filters: A specific one trained with frames from our video database (orange bars), and a generic one trained with the* fruits *dataset (blue bars). For all the videos analyzed (x-axis) the quality of the reconstruction is very similar with both dictionaries, showing our CSC-method does not require training a specific dictionary.*

**Choosing the best parameters**: We analyze the relative weight of each of the three terms in Eq. 8. We set $\beta_1$, which weights the sparsity term, to 10, and vary parameter $\beta_d$, which weights the data term, and parameter $\beta_2$, which weights the temporal smoothness term. Based on the MS-SSIM results shown in Figure 6, we choose for all our reconstructions the values $\beta_d = 100$ and $\beta_2 = 1$.

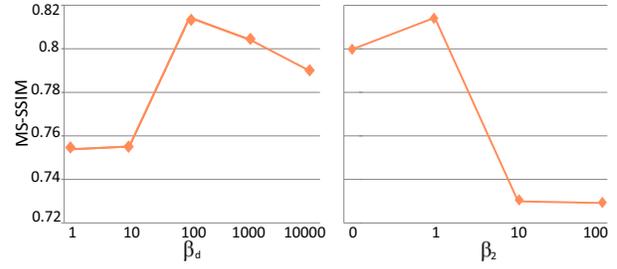

**Figure 6:** *Analysis of parameters $\beta_d$ and $\beta_2$ in Eq. 8, which control the weights of the data term and the temporal smoothness term, respectively. We plot the mean MS-SSIM value from eight reconstructed videos.*

## 7. Results and discussion

In this section we show and discuss our results with both sparse coding approaches: patch-based and convolutional. In recent work, Koller et al. [KSM*15] have performed an analysis of several state-of-the-art approaches for high spatio-temporal resolution video reconstruction with compressed sensing. They prove that the approach proposed by Liu et al. [LGH*13, HGG*11] achieves better reconstruction qualities than other state-of-the-art approaches. Therefore we compare the results of our convolutional sparse coding approach with our implementation of the framework proposed by Liu et al.

The parameters used in all the reconstructions are derived from the analysis in the previous section. The videos used for training in the patch-based approach are different from the reconstructed ones (see supplemental material), whereas for CSC we use an existing, generic dictionary trained from images of fruits [HHW15]. We have coded 20 frames in a single image. This number yields a good trade-off between quality and speed-up of the reconstructed video. For each frame, we sample less than 15% of the pixels. Despite this huge loss of information, we are able to reconstruct high-speed videos of good quality.

In general, both techniques yield results of similar quality, with a slight advantage for the patch-based approach in terms of MS-SSIM [WSB03] values, of about 0.05 on average (see also Figure 8). However, as discussed earlier in the paper, there are two important shortcomings of this approach: First, the need to train a dictionary made up of atoms containing similar structures to the reconstructed videos; otherwise the quality of the reconstruction degrades significantly. Second, training and reconstruction times are rather long (see Table 1 for reconstruction times). To overcome these, we introduced a second solution, based on convolutional sparse coding. This is not only significantly faster, but it also allows us to bypass the need to capture and train a dictionary, as discussed in Section 6.

As explained in the paper, a naïve 3D CSC approach quickly becomes computationally intractable, due to the huge convolutional matrices involved. On the other hand, reverting to a 2D, per-frame solution yields many artifacts in the results due to the low sampling

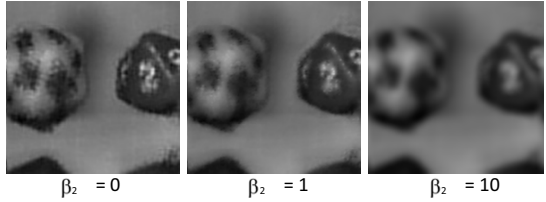

$\beta_2 = 0$ $\quad\quad\quad$ $\beta_2 = 1$ $\quad\quad\quad$ $\beta_2 = 10$

**Figure 7:** *Effect of our proposed temporal smoothness term on the reconstructed videos. From left to right: Result of CSC without this term ($\beta_2 = 0$), and two increasing values for the weight of the temporal smoothness term in the optimization $\beta_2$. Not including this term yields results with many artifacts due to the low sampling rate of each image separately (see also video in the supplemental material), while too high a value tends to over-blur the result.*

rate and the lack of temporal stability, as we show in Figure 7 (leftmost image) and the supplemental material. We therefore adapted the 2D convolutional sparse coding approach to our problem, taking advantage of the coded temporal information by enforcing sparsity of the derivatives in time while solving the optimization. Figure 7 (middle and right images) shows the effect of this term in the optimization, while Figures 9 and 10 show additional results with each of the two techniques (patch-based and CSC-based). Please refer to the supplemental material for the full videos.

Last, Table 1 shows reconstruction times for all the videos shown in this paper; on average, our CSC-based approach is 14x faster than a patch-based solution.

**Table 1:** *Reconstruction times for eight videos (20 frames each) with convolutional sparse coding and with patch-based sparse coding. Note the great speed-up achieved by the former.*

|           | Time (in seconds) ||
|-----------|---------------|-------------|
|           | Convolutional | Patch-based |
| Brain     | 249           | 4208        |
| Coke      | 239           | 3239        |
| Dice      | 236           | 3175        |
| Flower    | 237           | 3524        |
| Foreman   | 237           | 3746        |
| Balloon   | 236           | 3275        |
| FireHold  | 237           | 3071        |
| FireStart | 238           | 3180        |

## 8. Conclusions

Computational imaging aims at enhancing imaging technology by means of the co-design of optical elements and algorithms; capturing and displaying the full, high-dimensional plenoptic function is an open, challenging problem, for which compressive sensing and sparse coding techniques are already providing many useful solutions. In this paper we have focused on the particular case of high-speed video acquisition, and the intrinsic trade-off between temporal and spatial resolution imposed by bandwidth limitations. We have presented two sparse coding approaches, where we code the temporal information by sampling different time instants at every pixel. First, we have analyzed the key parameters in the patch-based sparse coding approach proposed by Liu et al. [LGH*13, HGG*11], which have allowed us to offer insights that lead to better quality in the reconstructed videos. We then have introduced a novel convolutional sparse coding framework, customized to enforce sparsity on the first-order derivatives in the temporal domain. The convolutional nature of the filter banks used in the reconstruction allowed for a more flexible and efficient approach, compared with its patch-based counterpart. We bypass the need to capture a database of high-speed videos and train a dictionary, while reconstruction times improve significantly.

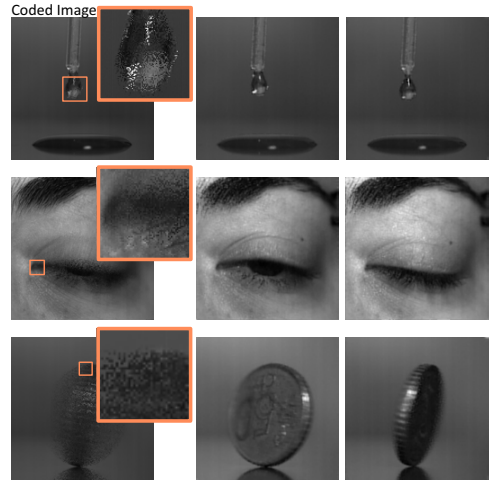

**Figure 9:** *Additional video sequences reconstructed using our CSC-based approach.* Left: *coded image which serves as input to the reconstruction algorithm; inset shows a close-up.* Right: *Two of the frames reconstructed for each sequence. Detail is recovered despite the large loss of information undergone during sampling. Note that the blur in the coin of the bottom row is not motion blur but due to limited depth of field instead.*

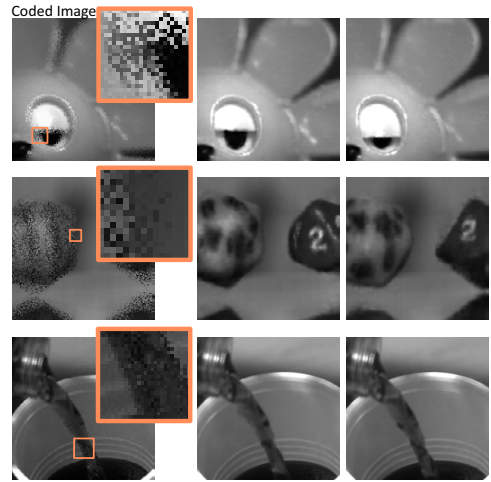

**Figure 10:** *Additional video sequences reconstructed using patch-based sparse reconstruction.* Left: *coded image which serves as input to the reconstruction algorithm; inset shows a close-up.* Right: *Two of the frames reconstructed for each sequence.*

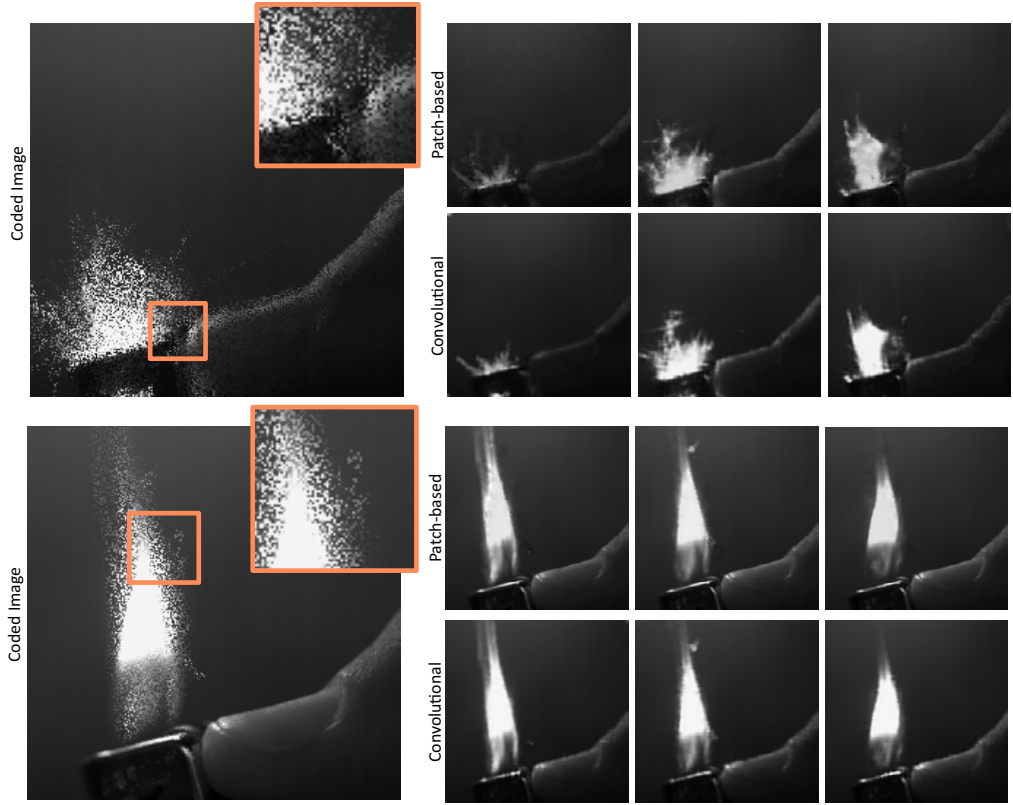

**Figure 8:** *Representative frames of two reconstructed videos:* FireStart *(top) and* FireHold *(bottom). The sequences show a lighter at different stages of ignition. We show reconstruction results for the two approaches discussed in the paper.* Left: *Input coded image, from which 20 frames will be reconstructed (inset shows a close-up).* Right: *Three sample frames of the reconstructed sequence, with the patch-based approach (top row) and with our CSC approach (bottom row).*

Many exciting venues for future research lie ahead. For instance, our strategy to impose an additional constraint in the temporal dimension is motivated by the fact that, due to the size of the convolutional matrices required to train the dictionary, it is not feasible to deal with (x-y-t) blocks directly. This is currently the main limitation of our approach, and it would be interesting to investigate other strategies in follow up work. Last, we hope that the development of computational techniques like ours will progressively allow the development of commercial imaging hardware with enhanced capabilities.

## 9. Acknowledgements

We would like to thank the Laser & Optical Technologies department from the Aragon Institute of Engineering Research (I3A), as well as the Universidad Rey Juan Carlos for providing a high-speed camera and some of the videos used in this paper. This research has been partially funded by an ERC Consolidator Grant (project CHAMELEON), and the Spanish Ministry of Economy and Competitiveness (projects LIGHTSLICE, LIGHTSPEED, BLINK, and IMAGER). Ana Serrano was supported by an FPI grant from the Spanish Ministry of Economy and Competitiveness; Elena Garces was partially supported by a grant from Gobierno de Aragon; Diego Gutierrez was additionally funded by a Google Faculty Research Award, and the BBVA Foundation; and Belen Masia was partially supported by the Max Planck Center for Visual Computing and Communication.

## References

[AEB06] AHARON M., ELAD M., BRUCKSTEIN A.: K-svd: An algorithm for designing overcomplete dictionaries for sparse representation. *IEEE Transactions on Signal Processing 54* (2006), 4311–4322. 4, 5

[AF13] ALMEIDA M. S. C., FIGUEIREDO M. A. T.: Frame-based image deblurring with unknown boundary conditions using the alternating direction method of multipliers. In *IEEE ICIP* (2013), pp. 582–585. 5

[BEL13] BRISTOW H., ERIKSSON A., LUCEY S.: Fast Convolutional Sparse Coding. In *Proc. CVPR* (2013), pp. 391–398. 2

[BL14] BRISTOW H., LUCEY S.: Optimization Methods for Convolutional Sparse Coding. In *arXiv:1406.2407* (2014). 2

[CPS*13] CHEN B., POLATKAN G., SAPIRO G., BLEI D., DUNSON D., CARIN L.: Deep learning with hierarchical convolutional factor analysis. *Pattern Analysis and Machine Intelligence, IEEE Transactions on 35*, 8 (Aug 2013), 1887–1901. 2

[CRT06] CANDES E., ROMBERG J., TAO T.: Robust uncertainty principles: exact signal reconstruction from highly incomplete frequency information. *IEEE Transactions on Information Theory 52* (2006), 489–509. 2

[Don06] DONOHO D.: Compressed sensing. *IEEE Transactions on Information Theory 52* (2006), 1289–1306. 2

[EHJT04] EFRON B., HASTIE T., JOHNSTONE I., TIBSHIRANI R.: Least angle regression. *Annals of statistics 32* (2004), 407–499. 5

[GBD*09] GUPTA A., BHAT P., DONTCHEVA M., CURLESS B., DEUSSEN O., COHEN M.: Enhancing and experiencing spacetime resolution with videos and stills. In *IEEE International Conference on Computational Photography* (2009). 2

[GHMN10] GU J., HITOMI Y., MITSUNAGA T., NAYAR S.: Coded rolling shutter photography: Flexible space-time sampling. In *IEEE International Conference on Computational Photography* (2010). 2

[GRKN07] GROSSE R. B., RAINA R., KWONG H., NG A.: Shift-invariant sparse coding for audio classification. In *Proceedings UAI* (2007), pp. 149–158. 2, 3

[HDL*14] HU X., DENG Y., LIN X., SUO J., DAI Q., BARSI C., RASKAR R.: Robust and accurate transient light transport decomposition via convolutional sparse coding. *Opt. Lett. 39*, 11 (Jun 2014), 3177–3180. 2

[HGG*11] HITOMI Y., GU J., GUPTA M., MITSUNAGA T., NAYAR S.: Video from a Single Coded Exposure Photograph using a Learned Over-Complete Dictionary. In *IEEE International Conference on Computer Vision (ICCV)* (2011), pp. 287–294. 2, 6, 7

[HHW15] HEIDE F., HEIDRICH W., WETZSTEIN G.: Fast and flexible convolutional sparse coding. In *Proc. CVPR* (2015). 2, 3, 4, 5, 6

[HST*14] HEIDE F., STEINBERGER M., TSAI Y.-T., ROUF M., PAJAK D., REDDY D., GALLO O., LIU J., HEIDRICH W., EGIAZARIAN K., KAUTZ J., PULLI K.: Flexisp: A flexible camera image processing framework. *ACM Transactions on Graphics 33*, 6 (2014). 2

[JCK16] JEON D. S., CHOI I., KIM M. H.: Multisampling Compressive Video Spectroscopy. *Computer Graphics Forum 35*, 2 (2016). 2

[KF14] KONG B., FOWLKES C. C.: *Fast Convolutional Sparse Coding (FCSC)*. Tech. rep., UC Irvine, 2014. 2

[KSM*15] KOLLER R., SCHMID L., MATSUDA N., NIEDERBERGER T., SPINOULAS L., COSSAIRT O., SCHUSTER G., KATSAGGELOS A. K.: High spatio-temporal resolution video with compressed sensing. *Opt. Express 23*, 12 (Jun 2015), 15992–16007. doi:10.1364/OE.23.015992. 6

[LFDF07] LEVIN A., FERGUS R., DURAND F., FREEMAN W. T.: Image and depth from a conventional camera with a coded aperture. *ACM Transactions on Graphics 26* (2007). 2

[LGH*13] LIU D., GU J., HITOMI Y., GUPTA M., MITSUNAGA T., NAYAR S.: Efficient Space-Time Sampling with Pixel-wise Coded Exposure for High Speed Imaging. *IEEE Transactions on Pattern Analysis and Machine Intelligence 36* (2013), 248–260. 2, 3, 4, 6, 7

[LLWD14] LIN X., LIU Y., WU J., DAI Q.: Spatial-spectral encoded compressive hyperspectral imaging. *ACM Transactions on Graphics 33* (2014), 1–11. 2

[MBPS10] MAIRAL J., BACH F., PONCE J., SAPIRO G.: Online learning for matrix factorization and sparse coding. *Journal of Machine Learning Research 11* (2010), 16–60. 4

[MCPG11] MASIA B., CORRALES A., PRESA L., GUTIERREZ D.: Coded apertures for defocus blurring. In *Ibero-American Symposium in Computer Graphics* (2011). 2

[MPCG12] MASIA B., PRESA L., CORRALES A., GUTIERREZ D.: Perceptually optimized coded apertures for defocus deblurring. In *Computer Graphics Forum* (2012), vol. 31, pp. 1867–1879. 2

[MWBR13] MARWAH K., WETZSTEIN G., BANDO Y., RASKAR R.: Compressive light field photography using overcomplete dictionaries and optimized projections. *ACM Transactions on Graphics 32* (2013), 1–11. 2

[MWDG13] MASIA B., WETZSTEIN G., DIDYK P., GUTIERREZ D.: A Survey on Computational Displays: Pushing the Boundaries of Optics, Computation, and Perception. *Computers & Graphics 37*, 8 (2013), 1012 – 1038. 2

[NM00] NAYAR S., MITSUNAGA T.: High dynamic range imaging: spatially varying pixel exposures. In *CVPR* (2000), vol. 1, pp. 472–479 vol.1. 2

[RAT06] RASKAR R., AGRAWAL A., TUMBLIN J.: Coded exposure photography: Motion deblurring using fluttered shutter. *ACM Transactions on Graphics 25* (2006), 795–804. 2

[RB11] RAVISHANKAR S., BRESLER Y.: Mr image reconstruction from highly undersampled k-space data by dictionaty learning. *IEEE Transactions on Medical Imaging 30* (2011), 1028–1041. 4

[SBN*12] SCHÖBERL M., BELZ A., NOWAK A., SEILER J., KAUP A., FOESSEL S.: Building a high dynamic range video sensor with spatially nonregular optical filtering. In *Proc. SPIE* (2012), vol. 8499, pp. 84990C–84990C–11. 2

[SGM15] SERRANO A., GUTIERREZ D., MASIA B.: Compressive high speed video acquisition. In *CEIG* (2015). 2

[SHG*16] SERRANO A., HEIDE F., GUTIERREZ D., WETZSTEIN G., MASIA B.: Convolutional sparse coding for high dynamic range imaging. *Computer Graphics Forum (Proc. EUROGRAPHICS) 35*, 2 (2016). 2, 4

[SKL10] SZLAM A., KAVUKCUOGLU K., LECUN Y.: Convolutional matching pursuit and dictionary training. *arXiv:1010.0422* (2010). 2

[VRA*07] VEERARAGHAVAN A., RASKAR R., AGRAWAL A., MOHAN A., TUMBLIN J.: Dappled photography: Mask enhanced cameras for heterodyned light fields and coded aperture refocusing. *ACM Trans. Graph. 26*, 3 (July 2007). 2

[WILH11] WETZSTEIN G., IHRKE I., LANMAN D., HEIDRICH W.: Computational Plenoptic Imaging. *Computer Graphics Forum 30*, 8 (2011), 2397–2426. 2

[WJV*04] WILBURN B., JOSHI N., VAISH V., LEVOY M., HOROWITZ M.: High-speed videography using a dense camera array. In *Computer Vision and Pattern Recognition* (2004), vol. 2, pp. 294–301. 2

[WLD*06] WAKIN M., LASKA J., DUARTE M., BARON D., SARVOTHAM S., TAKHAR D., KELLY K., BARANIUK R.: Compressive imaging for video representation and coding. In *Proceedings of the Picture Coding Symposium* (2006). 2

[WLGH12] WETZSTEIN G., LANMAN D., GUTIERREZ D., HIRSCH M.: Computational Displays. ACM SIGGRAPH Course Notes, 2012. 2

[WSB03] WANG Z., SIMONCELLI E., BOVIK A.: Multi-scale Structural Similarity for Image Quality Assessment. In *IEEE Conf. on Signals, Systems and Computers* (2003), pp. 1398–1402. 5, 6

[ZLN09] ZHOU C., LIN S., NAYAR S. K.: Coded Aperture Pairs for Depth from Defocus. In *IEEE International Conference on Computer Vision (ICCV)* (Oct 2009), pp. 325–332. 2